\documentclass[%
 aip,
 amsmath,amssymb,
 reprint,%
]{revtex4-1}

\usepackage{graphicx}
\usepackage{dcolumn}
\usepackage{bm}
\usepackage[utf8]{inputenc}
\usepackage[T1]{fontenc}
\usepackage{mathptmx}
\usepackage{etoolbox}
\usepackage{hyperref}

\makeatletter
\def\@email#1#2{%
 \endgroup
 \patchcmd{\titleblock@produce}
  {\frontmatter@RRAPformat}
  {\frontmatter@RRAPformat{\produce@RRAP{*#1\href{mailto:#2}{#2}}}\frontmatter@RRAPformat}
  {}{}
}%
\makeatother

\usepackage{amsmath,amsbsy}
\usepackage{url, color}

\begin{document}

\preprint{AIP/123-QED}
	
\title{Interplay of Turbulence and Proton-Microinstability Growth in Space Plasmas}	

\author{Riddhi Bandyopadhyay}
\affiliation{Department of Astrophysical Sciences, Princeton, NJ 08544, USA}
\email{riddhib@princeton.edu}
	
\author{Ramiz~A. Qudsi}
\affiliation{Center for Space Physics, Boston University, MA 02125, USA}

\author{S. Peter Gary}
\affiliation{Space Science Institute, Boulder, Colorado 80301, USA}
\thanks{deceased}

\author{William~H. Matthaeus}
\affiliation{Department of Physics and Astronomy, University of Delaware, Newark, DE 19716, USA}
\affiliation{Bartol Research Institute, Newark, DE 19716, USA}
	
\author{Tulasi~N. Parashar}
\affiliation{Department of Physics and Astronomy, University of Delaware, Newark, DE 19716, USA}
\affiliation{Bartol Research Institute, Newark, DE 19716, USA}

\author{Bennett~A. Maruca}
\affiliation{Department of Physics and Astronomy, University of Delaware, Newark, DE 19716, USA}
\affiliation{Bartol Research Institute, Newark, DE 19716, USA}

\author{Vadim Roytershteyn}
\affiliation{Space Science Institute, Boulder, Colorado 80301, USA}

\author{Alexandros Chasapis}
\affiliation{Laboratory for Atmospheric and Space Physics, University of Colorado Boulder, Boulder, CO 80303, USA}

\author{Barbara~L. Giles}
\affiliation{NASA Goddard Space Flight Center, Greenbelt, Maryland 20771, USA}

\author{Daniel~J. Gershman}
\affiliation{NASA Goddard Space Flight Center, Greenbelt, Maryland 20771, USA}

\author{Craig~J. Pollock}
\affiliation{Denali Scientific, Fairbanks, Alaska 99709, USA}

\author{Christopher~T. Russell}
\affiliation{University of California, Los Angeles, California 90095-1567, USA}

\author{Robert~J. Strangeway}
\affiliation{University of California, Los Angeles, California 90095-1567, USA}

\author{Roy~B. Torbert}
\affiliation{University of New Hampshire, Durham, New Hampshire 03824, USA}

\author{Thomas~E. Moore} 
\affiliation{NASA Goddard Space Flight Center, Greenbelt, Maryland 20771, USA}

\author{James~L. Burch}
\affiliation{Southwest Research Institute, San Antonio, Texas 78238-5166, USA}

\date{\today}

\begin{abstract}
Numerous prior studies have shown that as proton beta increases, a narrower range of proton temperature anisotropy values is observed. This effect has often been ascribed to the actions of kinetic microinstabilities because the distribution of observational data aligns with contours of constant instability growth rates in the beta-anisotropy plane. However, the linear Vlasov theory of instabilities assumes a uniform background in which perturbations grow. The established success of linear-microinstability theories suggests that the conditions in regions of extreme temperature anisotropy may remain uniform for a long enough time so that the instabilities have the chance to grow to sufficient amplitude. Turbulence, on the other hand, is intrinsically non-uniform and non-linear. Thin current sheets and other coherent structures generated in a turbulent plasma, may destroy the uniformity fast enough. It is therefore not a-priori obvious whether the presence of intermittency and coherent structures favors or disfavors instabilities. To address this question, we examined the statistical distribution of growth rates associated with proton temperature-anisotropy driven microinstabilities and local nonlinear time scales in turbulent plasmas. Linear growth rates are, on average, substantially less than the local nonlinear rates. However, at the regions of extreme values of temperature anisotropy, near the ``edges" of the populated part of the proton temperature anisotropy-parallel beta plane, the  instability growth rates are comparable or faster than the turbulence time scales. These results provide a possible answer to the question as to why the linear theory appears to work in limiting plasma excursions in anisotropy and plasma beta.
\end{abstract}

\maketitle

\section{Introduction} \label{sec:intro}
The interplanetary plasma  typically 
exhibits weak collisionality and strong turbulence~\citep{Bruno2005LRSP,Zimbardo2010SSR}. Similar conditions exist in many astrophysical systems. 
In such high-temperature, low-density magnetized plasmas, Coulomb collisions 
 between particles are rare, which allows the velocity distribution function (VDF) of a given particle species to persist in a state far from local thermodynamic equilibrium. Consequently the VDFs are generally non-Maxwellian, and
the distortions of the VDFs are manifested through substantial anisotropy in the pressure (or equivalently, temperature) tensor.
{Distorted VDFs can give rise to plasma microinstabilities \citep{Gary1993Book}
that can strongly influence dynamics, 
a topic that has received considerable recent attention
\citep{BaleEA09,MarucaEA11,Klein2018PRL}. 
However it is known that these distortions are themselves often caused, or amplified, by intermittency, in the form of coherent structures,
generated by nonlinearity in turbulent dynamics \citep{GrecoEA12-intermit,ServidioEA15}. 
Intermittency is also responsible for faster 
local turbulence time scales. 
This implies a competition between
linear and nonlinear processes,
necessarily involving 
enhancements of each due to 
intermittency. 
These effects have not been 
taken into account in 
previous 
treatments of the problem
\cite{BaleEA09,Klein2018PRL,Klein2019ApJ_Helios_revisit};
it is this competition that 
we address in the present paper.}

A simple VDF that exhibits temperature anistropy is the bi-Maxwellian, which has been extensively used is the plasma theory literature.  In this model, the VDF is 
characterized by 
well-defined
distinct temperatures $T_{\perp j}$ and $T_{\parallel j}$ 
for species $j$, referred 
to the directions perpendicular and parallel to the 
magnetic field.  The anisotropy of $j$-particles is then quantified by the ratio:
\begin{eqnarray}
R_{\mathrm{j}} = \frac{T^{\mathrm{j}}_{\perp}}{T^{\mathrm{j}}_{\parallel}} \label{eq:Rj}. 
\end{eqnarray}

Although  deviations from equilibrium are observed in all charged plasma species~\citep{Pilipp1987JGR,Maksimovic2005JGR,Marsch2006LRSP}, here we focus on protons. The extreme values of proton-temperature anisotropy in the solar wind exhibit a strong dependence on the parallel-proton beta~\citep{Gary2001GRL, Kasper2002GRL, HellingerEA06}
\begin{eqnarray}
\beta_{\parallel \mathrm{p}} = \frac{n_\mathrm{p}\,k_\mathrm{B}\,T^{\mathrm{p}}_{\parallel }}{B_0^2\,/\,(2\,\mu_0)} \ \label{eq:betap},
\end{eqnarray}
where, $n_\mathrm{p}$ is the proton number density, $k_\mathrm{B}$ is the Boltzmann constant, and $\mu_0$ is the permeability of vacuum. For progressively larger $\beta_{\parallel \mathrm{p}}$ values, the
range of observed temperature-anisotropy values narrows in the solar wind ~\citep{Kasper2002GRL} and the terrestrial magnetosheath~\citep{Maruca2018ApJ}. 

Kinetic microinstabilities \citep{Gary1993Book} offer an appealing theoretical explanation for the observed correlation between temperature anisotropy and plasma beta. Linearization of the Vlasov-Maxwell system about 
an assumed anisotropic equilibrium predicts that for extreme values of $R_\mathrm{p}$ and $\beta_{\parallel \mathrm{p}}$, the distribution function becomes unstable,
triggering the growth of 
waves. It is typically
assumed that upon reaching 
finite amplitude, these fluctuations
drive the plasma 
toward (temperature) isotropy.
The initial growth rate of the 
unstable waves is 
derivable via linear theory 
from the values of $\beta_{\parallel \mathrm{p}}$ and $R_\mathrm{p}$.

An important question 
is whether the unstable waves produced in this 
way 
are merely a passive ``side effect",
or if 
they actively modify the dynamics. Some authors adopt the interpretation that the ion-driven microinstabilities may ``feed" strong fluctuations 
\citep{Bale2005PRL}
in regions of 
instability, materially impacting the 
plasma dynamics.
{A 
different point of view
is that turbulence-cascade generated localized inhomogeneities, 
i.e, coherent 
structures such as 
current sheets
\citep{GrecoEA12-pre,ServidioEA15},
drive the temperature-anisotropies to extreme values, setting the stage for 
 linear instabilities
that might occur in regions of strong nonlinear effects.
Indeed recent studies such as 
{\it He et al.}~\cite{He2018ApJ_exhaust}
have shown in detail that velocity distributions are deformed by kinetic activity near intense current sheets, 
thus driving plasma 
activity such as the firehose instability,
which subsequently enhances turbulence.}   

The dissipation of turbulent fluctuations in weakly-collisional space plasmas involves the transfer of fluctuation energy from field and flow energies to thermal energies.  {The processes that contribute to this dissipation generally fall into one of two categories: strongly nonlinear intermittent processes well represented by particle-in-cell simulations and microinstability processes typically computed by linear dispersion theory in homogeneous plasmas.} Within the limited scope of hybrid kinetic simulaitons, where electrons are treated as neutralizing fluid, turbulence and microinstabilities have been shown to coexit~\citep{Kunz2014PRL,Markovskii2019ApJ,Hellinger2019ArXiv}.

Indeed, strong fluctuations 
are found near the same extreme regions of the $\beta_{\parallel \mathrm{p}}, R_\mathrm{p}$-plane where the instability growth rates are large, causing the plasma to remain (marginally) unstable to temperature-anisotropy instabilities~\citep{Bale2005PRL,Osman2012PRL,Servidio2014ApJ}. 
Similarly, computations of  shear-driven
turbulence \citep{Karimabadi2013PoP}
have shown that local  
instabilities can sporadically 
arise due to
 kinetic effects that 
are inevitably found near 
 current sheets and vortices
\citep{Greco2012PRE,ParasharMatthaeus16}.
From these studies, it is evident that regions contributing to strong
intermittency 
are also regions of 
strong
kinetic activity,
and furthermore these 
are often juxtaposed. It remains unclear which type of process -- linear or nonlinear-- dominates on average and determines the dynamics of large-scale phenomena. 
One may study 
this
relationship by comparing the relative time scales
of nonlinear and linear 
dynamical processes \citep{Matthaeus2014ApJ,Klein2018PRL}. There is some subtlety 
in this comparison when the medium is inhomogeneous, in that intermittency enters into this comparison in a significant way, while the standard instability calculation that we employ assumes extended plane wave solutions.

Recent studies of turbulence-driven cascade and temperature-anisotropy driven microinstability ~\citep{Matthaeus2014ApJ, Klein2018PRL, Hellinger2019ArXiv}
find that 
the 
majority of solar-wind intervals, in an idealized situation, would support the proton-driven microinstabilities.
However, the associated 
growth rates are rarely faster than all the other relevant time scales. Quantitatively, the non-linear time scales, estimated from the spectral amplitude near the ion-inertial scale, are faster than the growth rates for most of the analyzed samples. This comparison suggests that the turbulent cascade quickly destroys the ideal situation for harboring micro-instabilities which would, otherwise, grow to macroscopic values as 
unstable modes.

As suggested
above, 
the important physics of intermittency~\citep{Matthaeus2015PTRSA}
motivates modification of  
results obtained 
from globally based 
estimates such as \textit{average} non-linear time or \textit{average} spectral amplitude near a given scale.
Intermittent structures occupy a small fraction of the volume, but are likely 
responsible for 
a large fraction
of the plasma heating and particle energization~\citep{WanEA16}. Keeping this in mind, we propose that, instead of comparing 
timescales based on 
average fluctuation amplitude with 
growth rates, 
it is reasonable to 
compare the two 
based on 
the corresponding local values of plasma and turbulence
properties. 

To address the above issues, here
we carry out a local analysis of both 
the instability growth rates 
and the non-linear time scales. 
We analyze three datasets:
\begin{enumerate}
    \item A three-dimensional, kinetic, particle-in-cell (PIC) simulation,
    \item In situ observations of Earth's magnetosheath by the \textit{MMS} spacecraft, and
    \item In situ observation of the interplanetary solar wind by the \textit{Wind} spacecraft.
\end{enumerate}
For all three cases we will show that 
both instability growth rates and non-linear rates are intermittent with enhanced values near coherent structures~\citep{Osman2012PRL, Qudsi2020ApJ_instability}, 
and that, pointwise, the 
nonlinear processes are faster than the 
instabilities for a majority of cases. 

\section{Theory and Method} \label{sec:theory}
\textit{Linear Vlasov Theory}--
Solving the dispersion relation for the linearized Vlasov and Maxwell's equations in a homogeneous plasma, one obtains the 
angular frequencies, $\omega$, 
associated with a
given wavevector $\mathbf{k}$. The imaginary component of $\omega$ is the growth or decay rate of the $\mathbf{k}$ mode.
The dominant growth rate of a particular instability, 
expected in linear theory to trigger macroscopic effects, 
is:	
\begin{equation}
\gamma_{\max} \equiv \max_{\mathbf{k}} \Im(\omega) \ ,
\end{equation}	
where the maximum operation is taken over all wave-vectors 
$\mathbf{k}$ associated with that instability. The plasma is considered
unstable to a given instability if $\gamma_{\max} > 0$.

To calculate these growth rates, the technique and software of 
\cite{Maruca2012ApJ} and \cite{Maruca2018ApJ} are 
employed. For each pair of 
$(\beta_{\parallel {\mathrm{p}}},R_{\mathrm{p}})$-values, the value of $\gamma_{\max}$ is
determined for each of the four instabilities by computing the 
maximum value of $\Im(\omega)$ over a range of $\mathbf{k}$-values. For every point with $\gamma_{\max} > 0$, we select the maximum growth rate from the 4 types of instabilities, associated with proton-temperature 
anisotropy:
\begin{eqnarray}
\Gamma_{\mathrm{max}} = \max \{ \gamma_{\max}^{\mathrm{cyclotron}}, \gamma_{\max}^{\mathrm{mirror}} , \gamma_{\max}^{\mathrm{\parallel-firehose}}, \gamma_{\max}^{\mathrm{\nparallel-firehose}} \} \label{eq:gamma4},
\end{eqnarray}
{where each term inside the bracket represents the maximum growth rate for the corresponding instability, i.e., the cyclotron, mirror, parallel firehose and oblique firehose instabilities, respectively, as indicated by the superscript.} Values 
of $\Gamma_{\mathrm{max}}$ less than $10^{-5}\,\Omega_p$ are taken to be 
$0$ (i.e., effectively stable).
Note that in strong turbulence 
the plasma parameters vary significantly 
in space, so a separate calculation of $\Gamma_{\mathrm{max}}$ is required at each point $\bf r$. 
			
\textit{Nonlinear Timescales}--
The local nonlinear timescale, at a position $\mathbf{r}$, for a lengthscale $\ell$ can be estimated as
\begin{eqnarray}
\tau_{\mathrm{nl}} (\mathbf{r}) \sim {\ell}/{ \delta b_{\ell}} \label{eq:tnl}, 
\end{eqnarray}
where  the longitudinal magnetic field  
increment is 
\begin{eqnarray}
\delta b_{\ell} = \left \lvert\hat{\boldsymbol{\ell}} \mathbf{\cdot} \left[\mathbf{b} (\mathbf{r} + \boldsymbol{\ell}) - \mathbf{b} (\mathbf{r})\right]\right\lvert \label{eq:db},
\end{eqnarray}
and $\mathbf{b}$ is the \textit{total} magnetic field
expressed in Alfv\'en speed units. 
The vector lag $\boldsymbol{\ell}$ has a magnitude $\ell$ and direction $\hat{\boldsymbol{\ell}}$. 
The timescale 
$\tau_{\mathrm{nl}}({\bf r})$ is a strongly varying function of position, 
and may take on large values near coherent structures. {This local estimate of non-linear time is generalized from Kolmogorov's estimate of the average non-linear time scale which is given, at wavenumber $k$, by $\tau_{\mathrm{nl}}(k) = 1/(k u_k) \sim 1/(k \sqrt{E(k)})$.}
Accordingly, 
we  
compare 
the local values of $\gamma$ and $\tau_{\mathrm{nl}}$. 

For comparison with 
instability growth rates,
it is convenient to compute an equivalent frequency from the nonlinear timescales as
$\omega_{\mathrm{nl}} = 2 \pi / \tau_{\mathrm{nl}}$. Although majority of highly unstable modes are found to have the associate wavenumber close to the ion-inertial length ($\,d_{\mathrm{i}}$), to carefully evaluate the nonlinear frequency at the scale of the fastest growing mode and to account for the statistical spread in the associated wavenumber $k_{\mathrm{max}}$ we use a variable increment scale with 
\begin{eqnarray}
\ell = 1/k_{\mathrm{max}}\label{eq:kmax},
\end{eqnarray}
where $k_{\mathrm{max}}$ is the wavenumber associated with the fastest growing mode (separately for each point), as recently done by \citet{Klein2019ApJ_Helios_revisit}.

\section{PIC simulation}\label{sec:pic}
We analyze data obtained from 
a three-dimensional, fully kinetic, particle-in-cell (PIC) simulation~\citep{Roytershteyn2015PTRSA}. The simulation has $2048^3$ grid
points, with $L = 41.9\,d_{\mathrm{i}}$, $\beta_{\mathrm{p}}=\beta_{\mathrm{e}} = 0.5$, $m_{\mathrm{p}} / m_{\mathrm{e}} = 50$, $\delta B / B_0 = 1$. The
analysis is performed on a snapshot late in time evolution of the simulation. For more details, refer to \cite{Roytershteyn2015PTRSA}. We emphasize that 
no attempt is made to closely align the simulation parameters with those of the magnetosheath or the solar wind. {The three systems have very different plasma parameters, such as turbulence amplitude, plasma beta, and Reynolds number. A different initial condition in the simulation would have led to possibly different outcomes, but the three systems together cover plasma regimes rather broadly.}

\begin{figure}
	\begin{center}
		\includegraphics[width=0.8\linewidth]{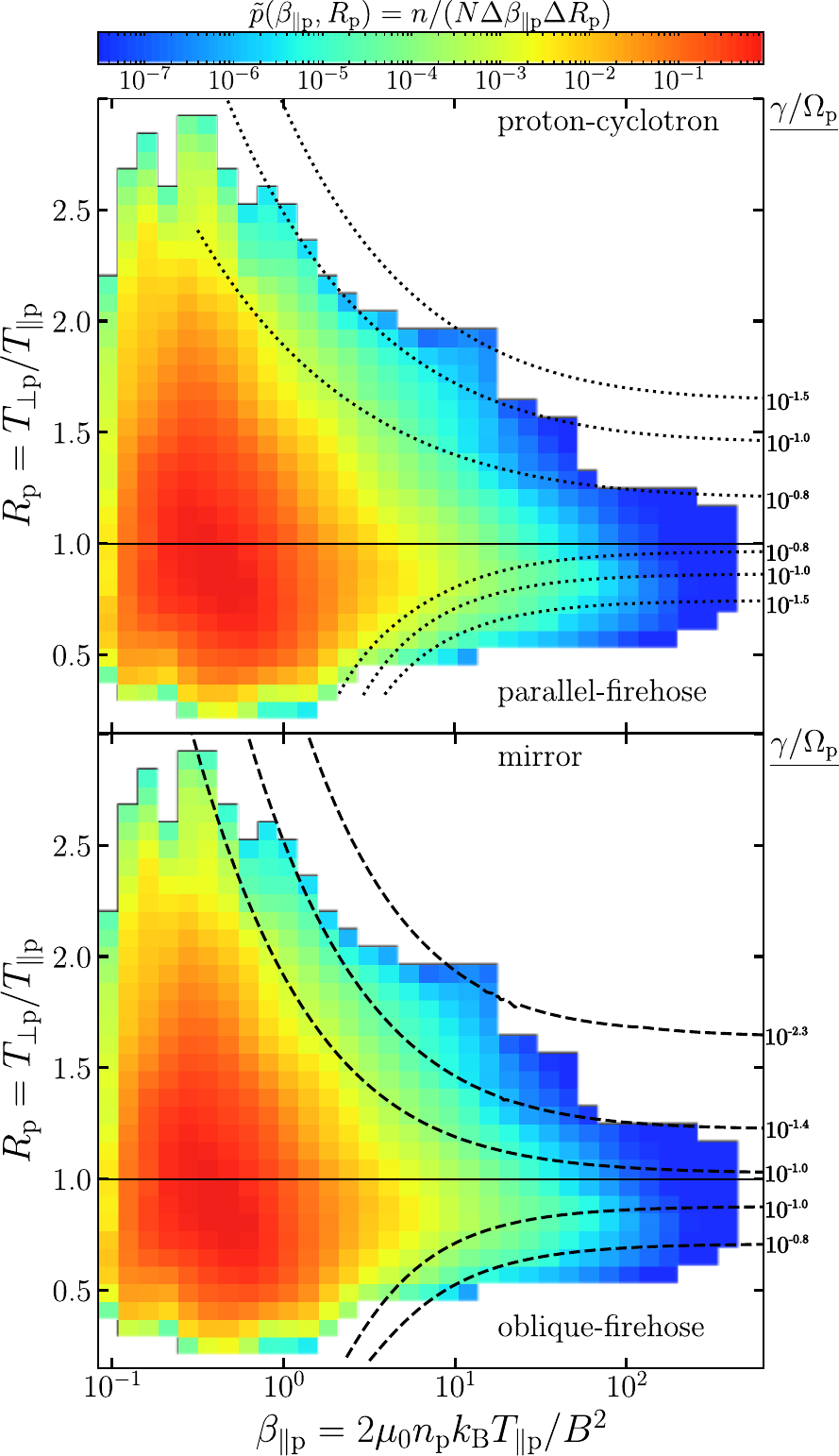}
		\caption{Two plots of the estimated probability density, $\tilde{p}$, of ($\beta_{\parallel \mathrm{p}}, R_\mathrm{p}$ )-values for the 3D PIC data. The two panels are identical
except for the overlaid curves, which show contours of constant growth rate for different instabilities. The curves in the top panel show the parallel instabilities: the proton-cyclotron ($R_\mathrm{p} > 1$) and parallel-firehose ($R_\mathrm{p} < 1$). The  curves in the bottom panel show the oblique instabilities: the mirror ($R_\mathrm{p} > 1$) and oblique-firehose ($R_\mathrm{p} < 1$). Each
contour is labeled with its growth rate, $\gamma$, in units of the proton cyclotron frequency, $\Omega_{\mathrm{p}}$.}
		\label{fig:brazil}
	\end{center}
\end{figure}

Figure~\ref{fig:brazil} shows the estimated values of probability density of $(\beta_{\parallel \mathrm{p}},R_{\mathrm{p}})$-values in the 3D PIC data
, along with the contours of constant instability growth rate,
indicating involvement 
of $\beta_{\parallel p}$-dependent constraints on 
$R_p$, in the simulation data. Although, for any given $\beta_{\parallel p}$-value, a distribution of $R_p$-values is observed, the distribution's {center} occurs near $R_p \approx 1$, and its width becomes progressively narrower with increased $\beta_{\parallel p}$.  Thus, the plasma likely hosts processes that favor isotropic proton-temperatures (limiting both $R_p>1$ and $R_p<1$) and these processes likely become more active at higher values of $\beta_{\parallel p}$. We believe these are the first reports of such  $\beta_{\parallel p}$-dependent constraints on $R_p$ in a three-dimensional, fully kinetic PIC simulation. Similar plots are obtained for the solar wind \citep{Maruca2012ApJ} and magnetosheath \citep{Maruca2018ApJ}.

\begin{figure}
	\begin{center}
		\includegraphics[width=1.05\columnwidth]{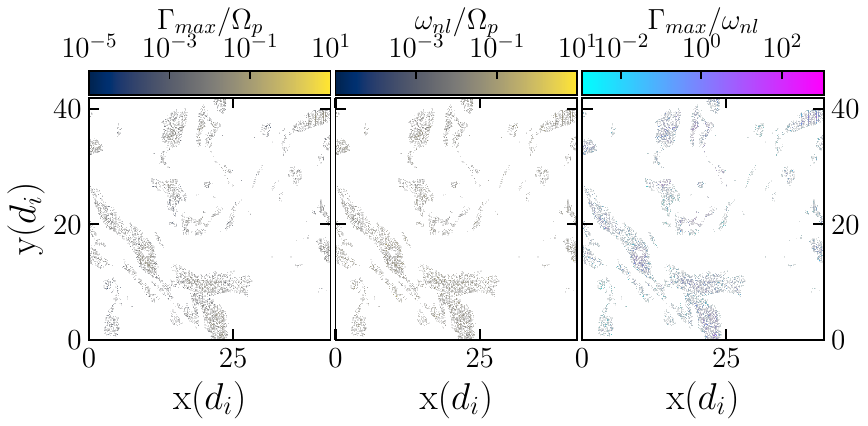}
		\caption{Plots (from left to right) of maximum growth rate $\Gamma_{\mathrm{max}}$, nonlinear frequency $\omega_{\mathrm{nl}}$ evaluated for a lag of  $\ell = 1/k_{\mathrm{max}}$, and the ratio $\Gamma_{\mathrm{max}} / \omega_{\mathrm{nl}}$ at $z \approx 35.6\,d_{\mathrm{i}}$ from PIC simulation.}
		\label{fig:pic}
	\end{center}
\end{figure}

The left panel of Fig.~\ref{fig:pic} shows the distribution of maximum growth rate, $\Gamma_{\mathrm{max}}$, (Eq.~\ref{eq:gamma4}) for a plane perpendicular to the mean magnetic field, at $z\approx35.6\,d_{\mathrm{i}}$.
The center panel illustrates the nonlinear frequencies at each point,  averaged over
lags of $\ell = 1/k_{\mathrm{max}}$ along the $x,y$, and $z$  directions. 
From the first two panels of Fig.~\ref{fig:pic}, it is evident that both kinds of frequencies are distributed 
intermittently in space, with clusters of 
large values in similar regions. 
However, from the right panel, the ratio of these frequencies rarely exceeds
unity. Even if both kind of processes are enhanced near the same regions of physical space, 
the non-linear processes are typically faster. Although Fig.~\ref{fig:pic} plots only one plane, later we show an analysis from the full 3D simulation domain.


\textit{In situ Observation}-- 
\begin{figure}
	\begin{center}
		\includegraphics[width=\linewidth]{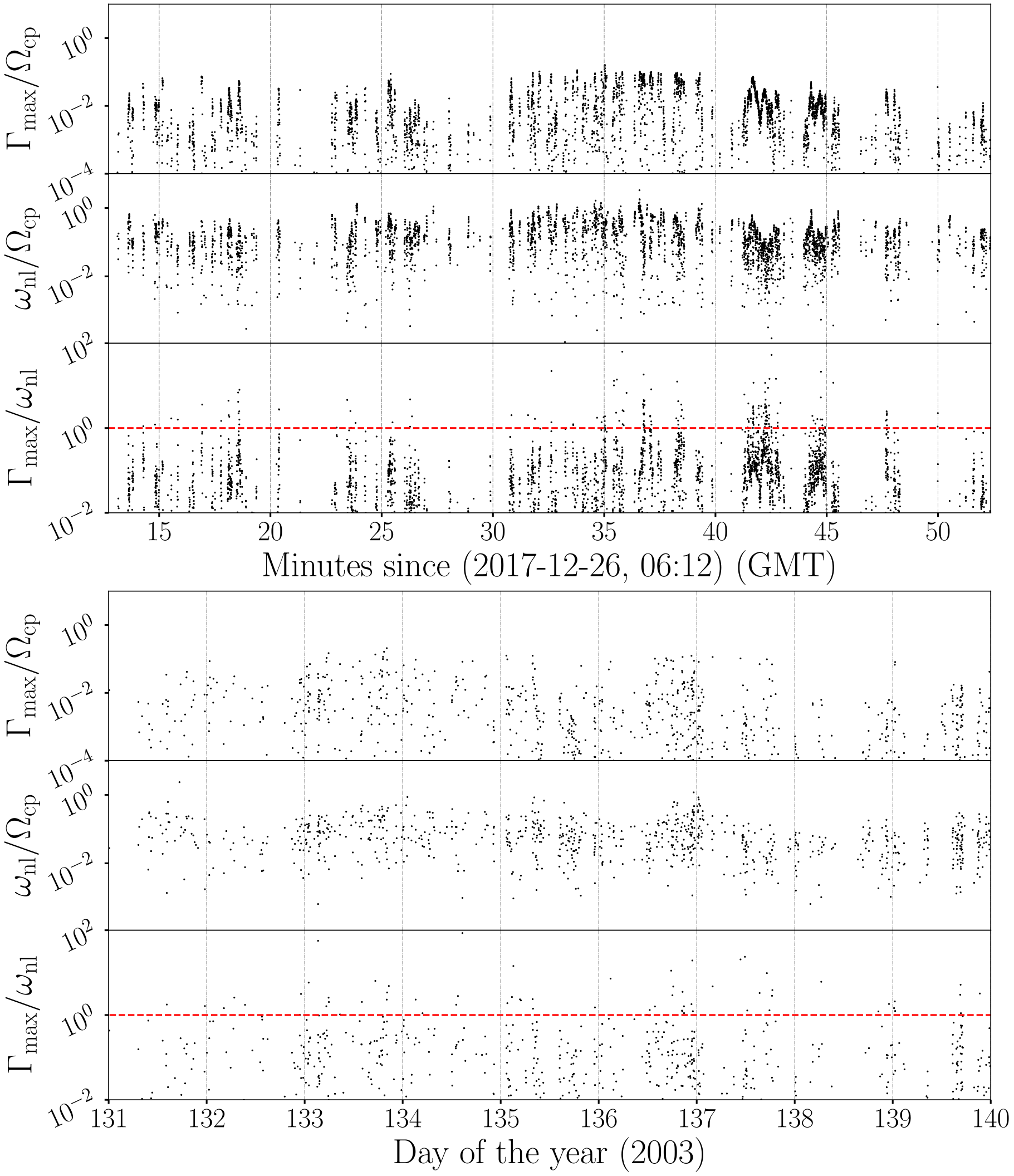}
		\caption{Time series of the maximum instability growth rates $\Gamma_{\mathrm{max}}$ (top), the nonlinear frequency $\omega_{\mathrm{nl}}$ at $\ell = 1/k_{\mathrm{max}}$ (middle), and the ratio $\Gamma_{\mathrm{max}} / \omega_{\mathrm{nl}}$ (bottom) for a burst-mode magnetosheath sample observed by the \textit{MMS} spacecraft (top) and an interplanetary solar wind interval sampled by the \textit{Wind} spacecraft (bottom). Note that due to the large difference in the measurement resolution of the \textit{MMS} and \textit{Wind} spacecraft, the time scales in the two figures are vastly different ($\sim 40$ min versus $\sim 10 $ days); however, they contain a similar number of correlation times of the respective data.}
		\label{fig:obs}
	\end{center}
\end{figure}
Though our analysis in the preceding section has important implications, the PIC simulation 
carries several limitations, e.g., artificial proton to electron mass ratio, small system size. Therefore, we next perform similar analyses, for two naturally occurring turbulent plasma systems: Earth's magnetosheath and the interplanetary solar wind. 

We use burst-mode \textit{MMS}~\citep{Burch2016SSR} data sampled in the Earth's magnetosheath for several burst-mode periods in both quasi-parallel and quasi-perpendicular shocked plasmas, including the ones reported in \cite{Maruca2018ApJ}. The \textit{MMS}/Fast Plasma Investigation~\citep{Pollock2016SSR} moments provide $\beta_{\parallel \mathrm{p}}$,  $R_{\mathrm{p}}$-values and magnetic-field measurements from the Flux Gate Magnetometer~\citep{Russell2016SSR} are used to compute the longitudinal increment (Eq.~\ref{eq:tnl}) at a spatial separation of $\ell = 1/k_{\mathrm{max}}$.
We select the magnetosheath intervals where the flow speed is greater than the Alfv\'en speed and use the  Taylor hypothesis to convert the temporal separation to  spatial separation $({\ell}= -\langle |\mathbf{V}| \rangle \tau)$. 
The non-linear frequencies were computed from the  
magnetic-field increments and interpolated to the ion cadence of 150 ms. The instability growth rates are calculated at ion cadence from the $\beta_{\parallel \mathrm{p}}$,  $R_{\mathrm{p}}$ values. 

{The specific intervals of MMS data that we analyze are the same as those reported in Reference \cite{Maruca2018ApJ}.}
The final statistics, shown later, are accumulated from all the intervals. However, in Fig.~\ref{fig:obs}, we show, as an example, a 40 min burst-mode sample from 06:12:43 - 06:52:23 UTC on 26 December 2017. Note that this interval is typical and not chosen for any special properties, other than the preliminary observation that it is turbulent and contains current sheets~\citep{Parashar2018PRL, Qudsi2020ApJ_instability}. The bottom panel on the top plot of Fig.~\ref{fig:obs} clearly shows that 
the ratio $\Gamma_{\mathrm{max}}/\omega_{\mathrm{nl}}$ for this interval rarely exceeds unity. {The fraction of 
points for which the instability rate is greater than the 
nonlinear rate are given in Table I.} 

In the bottom plot of Fig.~\ref{fig:obs}, we show a similar analysis for 1 au solar wind.
{The \textit{Wind} data used here are identical to those reported in \cite{Maruca2012ApJ}. }
We use measurements from \textit{Wind} satellite, accumulated over a period of about 10 {days}. We use  $11\,\mathrm{Hz}$ magnetic field measurements from \textit{Wind}'s Magnetic Field Investigation~\citep{Lepping1995SSR} to calculate $\omega_{\mathrm{nl}}$ for a Taylor-shifted separation of $\ell = 1/k_{\mathrm{max}}$. The two Faraday cups in the Solar Wind Experiment~\citep{Ogilvie1995SSR} return one ion spectrum every $\approx 90\,\mathrm{s}$ and the $\omega_{\mathrm{nl}}$ values are interpolated to this cadence. A bi-Maxwellian distribution is fit to each ion spectrum to compute proton moments~\cite{MarucaKasper13} and thus infer values of $R_{\mathrm{p}}$ and $\beta_{\parallel \mathrm{p}}$. In the small sample of $\approx 10$ days of \textit{Wind} data, shown in the bottom plot of Fig.~\ref{fig:obs}, the exhibited behavior closely resembles the magnetosheath results (Fig.~\ref{fig:obs}, top), apart from the differences in time scales. Again, the nonlinear frequency, $\omega_{\mathrm{nl}}$, is greater than the instability growth rate, $\gamma$, for the majority, and the regions in which the growth rate is of
relative significance are 
sporadic.

\begin{table}
\caption{Fraction of points with $\Gamma_{\mathrm{max}}
  >\omega_{\mathrm{nl}}$ for different systems.
} 
\label{tab:ratio}
\begin{tabular}{c | c | c | c }
 & 3DPIC & MMS & Wind \\ \hline
Number of points with $\Gamma_{\mathrm{max}}/\omega_{\mathrm{nl}} < 1$ & 11$\%$ & 4$\%$ & 12$\%$ \\
\hline
\end{tabular}
\end{table}

\begin{figure}
	\begin{center}
		\includegraphics[width=\linewidth]{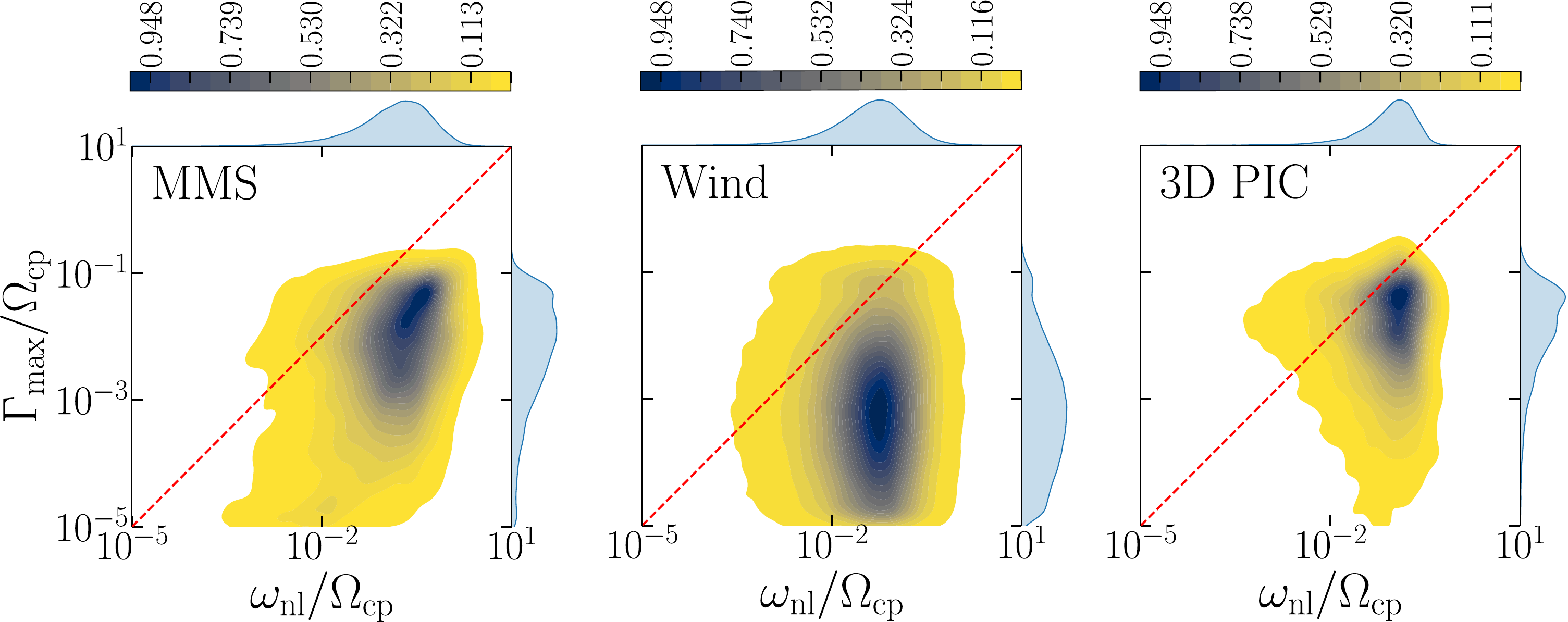}
		\caption{Joint probability distribution functions of the maximum instability growth rate $\Gamma_{\mathrm{max}}$ and the nonlinear frequency $\omega_{\mathrm{nl}}$ from PIC simulation, \textit{MMS} data in the magnetosheath, and \textit{Wind} data in the interplanetary solar wind.  The fraction of points above the 
		$\Gamma_{\mathrm{max}} =\omega_{\mathrm{nl}}$ line is about 4$\%$ for the MMS data, 12 $\%$ for the Wind data, and 11$\%$ for the 3D PIC data.  Only $\gamma>0$ cases are considered here. This is further discussed in Section~\ref{sec:conc}.}
		\label{fig:kde}
	\end{center}
\end{figure}

A key result of this paper is shown in Fig.~\ref{fig:kde}. Here, we plot joint probability distribution functions of the instability growth rates $(\Gamma_{\mathrm{max}})$ and the non-linear frequencies $(\omega_{\mathrm{nl}})$ for all three datasets. These plots represent a concise but compelling way of comparing linear theory results against nonlinear results, while at the same time enabling a comparison of observations versus simulations.
Note that all the rates are normalized by the respective cyclotron frequency, allowing direct comparison.
In all three cases, the core of the distribution resides well below the $\Gamma_{\mathrm{max}} = \omega_{\mathrm{nl}}$ line. 
From this result, we see that
for most data samples, 
the non-linear processes are faster than the linear-instability growth, 
when the substantial nonuniformity, or intermittency, of both types of processes are taken into account.

\begin{figure}
	\begin{center}
		\includegraphics[width=\linewidth]{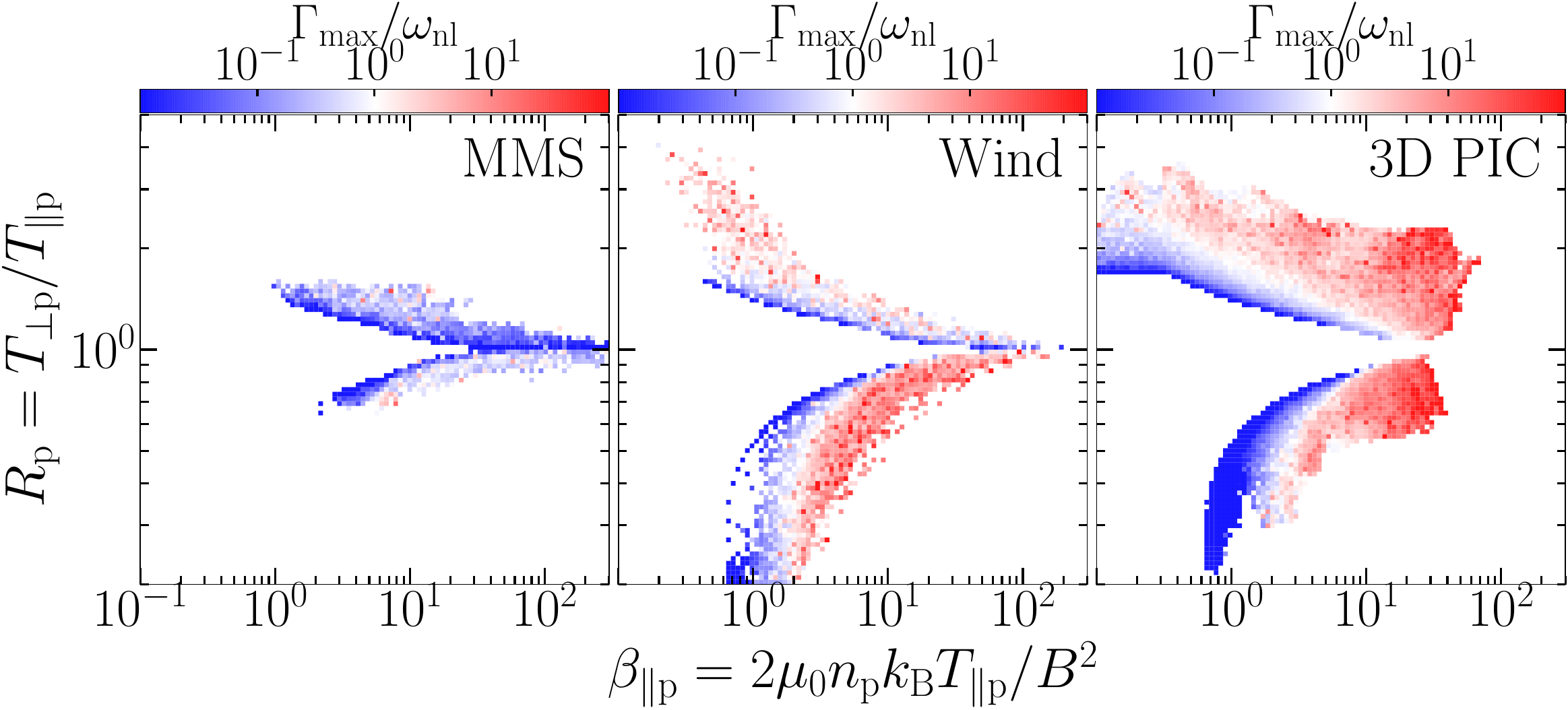}
		\caption{Distribution of the ratio of the maximum linear growth rate ($\Gamma_{\mathrm{max}}$) to the nonlinear frequency ($\omega_{\mathrm{nl}}$) evaluated at the wavenumber of maximum growth rate on the ($\beta_{\parallel \mathrm{p}}, R_\mathrm{p}$ )-plane for the three cases.}
		\label{fig:ratio}
	\end{center}
\end{figure}

{All of the datasets show that non-linear time scales are in general faster than the fastest of the linear timescales. This suggests that in most cases the linear instabilities do not have enough time to grow in the plasma in a way that is significant enough to affect the dynamics or the statistical behaviour of whole plasma.

On the other hand, decades of studies indicate that linear theory is effective in 
predicting the boundaries of ($\beta_{\parallel \mathrm{p}}, R_\mathrm{p}$)-plots which suggest that linear instabilities regulate extreme values of $R_\mathrm{p}$. 
To explore this, we look at the distribution of the two frequencies ($\Gamma_{\mathrm{max}}$ and $\omega_{\mathrm{nl}}$) and their ratio for the three datasets (PIC, MMS, and Wind) on the ($\beta_{\parallel \mathrm{p}}, R_\mathrm{p}$)-plane. Fig.~\ref{fig:ratio} shows this ratio in the three cases. Here we see that the region along the edges, which is most susceptible to instability is where points with $\Gamma_{max}/\omega_{nl} > 0.1$ lie. For all the three cases, ratio of the two frequencies is increasing as we move outside from the centroid of the distribution. In all three cases the linear time scales become comparable or faster than their non-linear counterpart near the edges of the ($\beta_{\parallel \mathrm{p}}, R_\mathrm{p}$)-plots. {This result shows that although in majority of the plasma $\Gamma_{\mathrm{max}}$ is less than $\omega_{\mathrm{nl}}$, it is greater than $\omega_{\mathrm{nl}}$ along the periphery of the populated part of the ($\beta_{\parallel \mathrm{p}}, R_\mathrm{p}$)-plane.}
Consideration of this pattern 
may explain how
the instabilities are quite efficient at limiting the extension 
of the plasma population to extreme anisotropy regions.}

{We note that the boundaries from $\Gamma_{\mathrm{max}}$/$\omega_{\mathrm{nl}}<1$ to $\Gamma_{\mathrm{max}}$/$\omega_{\mathrm{nl}}>1$ in the ($\beta_{\parallel \mathrm{p}}, R_\mathrm{p}$)-plane are very similar in all three panels. These critical lines signify the effectiveness of non-linear versus linear processes in the plasmas. The similarity in the lines in all the three graphs suggests that the transition occurs in a similar manner in all cases.
The pattern of the MMS data is notably less extended than the other two cases. One possibile explaination of this is the generally greater level of turbulence in the magnetosheath, as well as other potential factors such as, including factors such as degree of compressibility, differing system size, nature of driving. Exploration of these differences is beyond the current scope of the study. 
}

\section{Discussion \& Conclusions} \label{sec:conc}
Temperature-anisotropy driven microinstabilities are often considered to constrain the temperature anisotropy values in weakly-collisional plasmas~\citep{Kasper2002GRL,Hellinger2006GRL,Verscharen2019LRSP}. Recall that the linear Vlasov theory of  instabilities assumes a homogeneous background, in which background a small perturbation grows exponentially. The established success of linear-microinstability theories suggests that the conditions near the extremely anisotropic temperature may be uniform enough to justify an application of homogeneous linear theories. Turbulence, on the other hand, is intrinsically non-uniform and nonlinear. Thin current sheets, and other coherent 
structures generated
by the energy cascade, are sites of extreme temperature anisotropy \cite{Greco2012PRE}
and therefore, the high growth rates due to the microinstabilities also reside in the same vicinity. It is therefore not 
a priori obvious whether the presence of intermittency and 
coherent structures favors or disfavors instabilities 
in comparison with 
nonlinear effects.
This question
has motivated the present study.

To address this question, 
we have examined the 
statistical distribution of 
growth rates associated with proton temperature-anisotropy driven microinstabilities and the local nonlinear time scales,
in three distinct systems. The three systems cover different ranges of ($R_{\mathrm{p}}, \beta_{\parallel \mathrm{p}}$)-values among other parameters. However, both simulation and observation results show that, when the comparison is performed in this way, locally in space, 
only a small fraction of the samples support long-lived  linear instabilities.
Naively, one may conclude that the nonlinear effects
do not allow sufficient time for the instabilities to grow large enough to affect the global dynamics to any significant degree (Fig.~\ref{fig:kde}). {Yet, decades of observations present strong evidence that linear microinstabilities regulate ion temperature anisotropy. A number of studies using in-situ observations~\citep[e.g.,][]{Gary2001GRL, Kasper2002GRL, Hellinger2006GRL, Maruca2012ApJ, Maruca2018ApJ}, have found that the distribution of plasma over the ($R_{\mathrm{p}}, \beta_{\parallel \mathrm{p}}$)-plane is well restricted by the thresholds predicted by linear Vlasov theory. The results shown in Fig.~\ref{fig:brazil} provide further evidence.

Fig.~\ref{fig:ratio} presents a partial resolution of this apparent contradiction between linear and non-linear processes. We see that although $\omega_{\mathrm{nl}} > \Gamma_{\mathrm{max}}$ for most data, linear instabilities do become faster and possibly even disrupt the turbulence cascade at most extreme anisotropies, for which microkinetic instabilities are expected to be most active.

{Klein et al. 2018~\cite{Klein2018PRL} find that about 50$\%$ of all cases are unstable, and less than 10$\%$ of those cases have growth rates
faster than the non-linear time. We find that the fraction of data points with $\gamma >0$ is about 28$\%$ for the MMS data, and 16 $\%$ for the Wind data., and 16$\%$ for the 3D PIC data, 
Within the set of unstable points about 4$\%$ for the MMS data and 12 $\%$ for the Wind data, and 11$\%$ for the 3D PIC data 
show $\Gamma_{\mathrm{max}} > \omega_{\mathrm{nl}}$. These differences are likely due to our use of a {\it local} non-linear time as well as our approximation of the VDFs as bi-Maxwellian.}

These illuminating results motivate further studies that incorporate more sophisticated analyses.  In particular, the present study utilized several simplifying assumptions.  This work has assumed that ion VDFs are well modeled as bi-Maxwellians.  Though this is often the case for the ``core'' portion, the full VDF frequently exhibits a ``beam'' and/or other features.  Each of these
additional non-Maxwellian features can 
modify the free energy available 
to drive instabilities, thus
modifying the computed growth rates~\citep[e.g.,][]{Martinovic2021ApJ_Helios}. {Further, the fluctuation growth rates or dissipation rates estimated by linear theory are often less than that measured in practice~\citep[see][]{He2019ApJ_direct, He2020ApJ_spectra}.
The approach suggested by He et al \cite{He2019ApJ_direct, He2020ApJ_spectra} might lead to a greater range of applicability of linear theory, associated with taking into account many unstable waves. 
}
Including such 
refinements would represent a significant extension of the present 
study.}

\begin{acknowledgments}

We dedicate this article to the memory of our co-author S. Peter Gary. Throughout his illustrious career-- which included literally writing the book on microinstabilities in space plasmas-- he remained a tireless, insightful collaborator and a patient, supportive mentor to many in the field.

TNP was supported by NSF SHINE Grant AGS-1460130 and NASA HGI Grant 
	80NSSC19K0284. WHM is a member of the \textit{MMS} Theory and Modeling Team and was 
	supported by NASA Grant NNX14AC39G. The research of SPG was supported by 
	NASA grant NNX17AH87G.
	
The simulation described in this paper was performed as a part of the Blue Waters sustained-petascale computing project, which was supported by the National Science Foundation (awards OCI-0725070 and ACI-1238993) and the State of Illinois. Blue Waters allocation was provided by NSF PRAC award 1614664.

The authors would like to thank Rohit Chhiber for useful discussions and Yan Yang for helping to prepare the PIC dataset. 
\end{acknowledgments}

\section*{Data Availability}\label{sec:data}
This study used Level 2 FPI and FIELDS data according to the guidelines 
	set forth by the \textit{MMS} instrumentation team. All data are freely available at 
	\url{https://lasp.colorado.edu/MMS/sdc/}, Refs. \cite{Burch2016SSR, Russell2016SSR, Pollock2016SSR}. We thank the \textit{MMS} SDC, FPI, and FIELDS 
	teams for their assistance with this study. \\
	\textit{Wind} SWE and MFI data are available from CDAWeb (\url{https://cdaweb.gsfc.nasa.gov/}), Refs. \cite{Lepping1995SSR, Ogilvie1995SSR}.  The authors thank the \textit{Wind} team for the \textit{Wind} magnetic field and plasma data.
	
	The bi-Maxwellian analysis code is summarized in \cite{MarucaKasper13}.
	
\section*{References}\label{sec:ref}

%

\end{document}